\documentclass[10pt]{article}
\usepackage{graphicx}
\pdfoutput=1

\def\Title#1{\begin{center} {\Large #1 } \end{center}}
\def\Author#1{\begin{center}{ \sc #1} \end{center}}
\def\Address#1{\begin{center}{ \it #1} \end{center}}

\newcommand\pubblock{\rightline{\begin{tabular}{l} Proceedings of the Second Annual LHCP\\ \pubnumber\\
         \pubdate  \end{tabular}}}

\newenvironment{Abstract}{\begin{quotation} \begin{center} 
             \large ABSTRACT \end{center}\bigskip 
      \begin{center}\begin{large}}{\end{large}\end{center} \end{quotation}}

\newenvironment{Presented}{\begin{quotation} \begin{center} 
             PRESENTED AT\end{center}\bigskip 
      \begin{center}\begin{large}}{\end{large}\end{center} \end{quotation}}

\def\Acknowledgements{\bigskip  \bigskip \begin{center} \begin{large}
             \bf ACKNOWLEDGEMENTS \end{large}\end{center}}

\def\beq{\begin{equation}}
\def\eeq#1{\label{#1}\end{equation}}
\def\eeqn{\end{equation}}

\def\beqa{\begin{eqnarray}}
\def\eeqa#1{\label{#1}\end{eqnarray}}
\def\eeqan{\end{eqnarray}}

\let\bar=\overbar

\def\Dslash{\not{\hbox{\kern-4pt $D$}}}
\def\dslash{\not{\hbox{\kern-2pt $\del$}}}

\newcommand{\eslash}{{\hbox{$E$\kern-0.6em\lower-.05ex\hbox{/}\kern0.10em}}}
\newcommand{\hslash}{{\hbox{$H$\kern-0.8em\lower-.05ex\hbox{/}\kern0.10em}}}

\def\msb{{\bar{\ssstyle M \kern -1pt S}}}

\newcommand{\zjets}{\ensuremath{Z(\nu\bar{\nu})}+jets\xspace}
\newcommand{\ttbar}{\ensuremath{t\bar{t}}\xspace}
\newcommand{\tev}{TeV\xspace}
\newcommand{\gev}{GeV\xspace}
\newcommand{\fb}{\ensuremath{\mbox{fb}^{-1}}\xspace}
\newcommand{\sqs}{\ensuremath{\sqrt{s}}\xspace}
\newcommand{\mht}{\ensuremath{\hslash_T}\xspace}
\newcommand{\met}{\ensuremath{\eslash_T}\xspace}
\newcommand{\njet}{\ensuremath{n_{jet}}\xspace}
\newcommand{\nb}{\ensuremath{n_b}\xspace}
\newcommand{\Ht}{\ensuremath{H_T}\xspace}
\newcommand{\mttwo}{\ensuremath{M_{T2}}\xspace}
\newcommand{\dphiwl}{\ensuremath{\Delta\phi(\mbox{W},\ell)}\xspace}
\newcommand{\stlep}{\ensuremath{S_T^{\ell}}\xspace}
\newcommand{\Pt}{\ensuremath{p_T}\xspace}

\newcommand{\RNum}[1]{\uppercase\expandafter{\romannumeral #1\relax}}

\def\neu#1{\widetilde\chi^0_{#1}}

\def\stop#1{\ensuremath{\tilde{t}_{#1}}}

\newcommand{\tonetttt}{\ensuremath{\tilde{g} \rightarrow t\bar{t}\neu1 \xspace}}
\newcommand{\tonetonet}{\ensuremath{\tilde{g} \rightarrow \tilde{t} \bar{t} \rightarrow t\bar{t}\neu1}}
\newcommand{\tone}{\ensuremath{\tilde{g} \rightarrow q\bar{q}\neu1 \xspace}}
\newcommand{\ttwo}{\ensuremath{\tilde{q} \rightarrow q\neu1 \xspace}}
\newcommand{\ttwobb}{\ensuremath{\tilde{b} \rightarrow b\neu1 \xspace}}

\usepackage{color}
\usepackage{xspace}
\usepackage{caption}
\usepackage{subcaption}
\usepackage{mathtools}

\newcommand\pubnumber{ CMS CR-2014/203 }

\newcommand\pubdate{\today}

\def\affiliation{
On behalf of the CMS Experiment, \\
School of Physics \\
University of Bristol, UK. }

\begin{document}

\large
\begin{titlepage}
\pubblock

\vfill
\Title{  Inclusive searches for SUSY at CMS  }
\vfill

\Author{ Chris Lucas }
\Address{\affiliation}
\vfill
\begin{Abstract}

Multiple searches for supersymmetry have been performed at the CMS 
experiment. Of these, inclusive searches aim to remain as sensitive as 
possible to the widest range of potential new physics scenarios.
The results presented in this talk use the latest 19.5\fb of 8 \tev 
data from the 2012 LHC run. Interpretations are given within the context of 
Simplified Model Spectra for a variety of both hadronic and leptonic 
signatures.
\end{Abstract}	
\vfill

\begin{Presented}
The Second Annual Conference\\
 on Large Hadron Collider Physics \\
Columbia University, New York, U.S.A \\ 
June 2-7, 2014
\end{Presented}
\vfill
\end{titlepage}
\def\thefootnote{\fnsymbol{footnote}}
\setcounter{footnote}{0}
%

\normalsize 


\section{Introduction}

Of the many proposed beyond the Standard Model (SM) theories, Supersymmetry (SUSY)
still remains one of the best theoretically motivated and studied.
A low energy realisation of SUSY with TeV-scale
third-generation squarks \cite{Barbieri:2009ev} is motivated by the cancellation of the quadratically divergent
loop corrections to the mass of the recently discovered Higgs boson \cite{Aad:2012tfa,Chatrchyan:2012ufa} in the SM without
the need for significant fine tuning. For R-parity conserving SUSY \cite{Farrar:1978xj}, supersymmetric particles
(sparticles) such as squarks and gluinos are produced in pairs and decay to the lightest,
stable supersymmetric particle (LSP). The LSP is generally assumed to be weakly interacting
and massive, hence a typical signature is a final state of multijets accompanied by significant
missing transverse energy, \met. Such a weakly interacting massive particle (WIMP) is considered
to be a prime candidate for Dark Matter (DM).

Traditionally searches have interpreted results in terms of complete phenomenological models such as the Constrained Minimally Supersymmetric extension to the Standard Model (CMSSM)\cite{Chatrchyan:2011zy}. More recently however,
analyses have moved from model-specific to signature-specific interpretations. Simple models are constructed which contain pair-produced
sparticles which each decay with a 100\% branching fraction to a given final state. By varying the sparticle masses, model scans and interpretations are produced which can be a applied to a wide range of phenomenological models.
Such models are known are Simplified Model Spectra (SMS) and will be the focus of the interpretations in the presented results.

CMS SUSY searches cover an expansive amount of SUSY decay phase space through the multiple, 
complimentary search regions and methodologies used.
Perhaps the most wide-reaching are so-called `inclusive searches' which aim to 
remain as generic as possible.

\section{Analyses}	

In these proceedings multiple complimentary searches using the CMS detector \cite{Chatrchyan:2008aa} are detailed. Both hadronic 
and leptonic
analyses are covered, with results and interpretations given in terms of 
a variety of different Simplified Model Spectra (SMS) models \cite{Alwall:2008ag}. All of the searches detailed use the full CMS 2012 
dataset consisting of 19.5\fb of data at \sqs=8\tev, applying the latest CMS
recommended reconstruction and ID criteria. Despite the different search 
techniques employed, each analysis has the common technique of utilising the 
large amounts of missing transverse energy (\met) due to sparticle decays and 
the subsequent, non-interacting LSP.

Common kinematic variables used in the following analyses include the scalar sum of hadronic event activity, \Ht, the vectorial sum of hadronic event activity, \mht and the previously mentioned vectorial sum of the transverse energies of all particles, \met.

\subsection{Multijet + \met}
\label{sec:multijetmet}
A search for SUSY is performed in the all-hadronic channel, searching for large 
jet multiplicities and significant \met, while vetoing any leptonic activity in 
the event \cite{Chatrchyan:2014lfa}. This particular analysis remains inclusive by making no specific 
requirement on the number of b-tagged jets in the final state.
The dominant backgrounds for such a hadronic analysis can be divided 
into sources of genuine and fake \met, each of which will be described in the 
following paragraphs.

Genuine sources of \met involve a decay of some 
particle to a final state containing at least one neutrino. The largest of these 
is from Z boson production, where the Z decays to a neutrino pair, with 
associated jet production. Similarly, 
decays of a W boson, whether directly produced or via top quark decay, produce a 
lepton and a neutrino, and are therefore able to enter the hadronic signal region of such 
an analysis when the lepton is either mis-identified or missed entirely. 
The background contribution from \zjets is estimated using a 
photon-enriched data control sample, relying on the kinematic similarity between Z and 
$\gamma$ bosons at high-\Pt. In order to mimic the \met from the neutrinos in 
the Z-decay, the $\gamma$ is not included in the calculation of 
discriminating event variables such as \Ht and \mht. The ratio of the two 
production cross-sections ($R_{Z/\gamma}$) is measured in both simulation and 
data, before calculating a prediction which can be extrapolated into the signal 
region. Similar data-driven techniques are applied to estimate lost-lepton 
contributions.

The main source of fake \met comes from QCD multijet events, where jet
mis-measurements can lead to significant amounts of \met coupled with large 
numbers of jets. This background contribution is estimated using a rebalance and 
smearing technique on the event's jets \cite{Chatrchyan:2014lfa}, performed in kinematic sidebands to the signal region.

\begin{figure}[t]
\centering
\includegraphics[width=.7\textwidth]{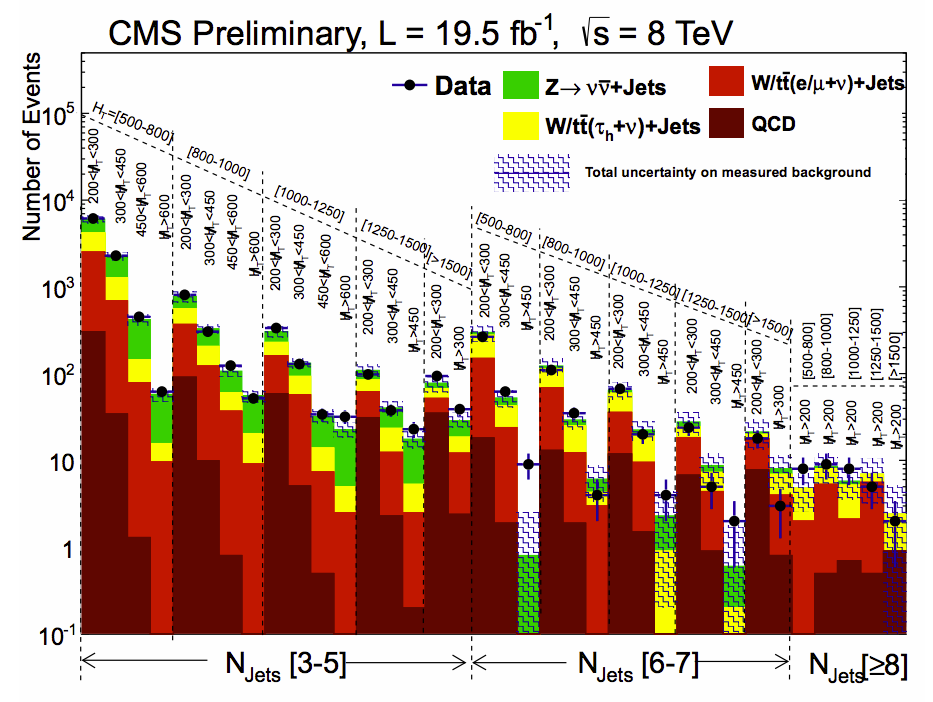}
\caption{Data observations compared with background predictions for the Multijet + \met
 analysis, split into bins of \njet, \Ht and \mht.}
\label{fig:mjet_results}
\end{figure}

\begin{figure}[b!]
        \centering
        \begin{subfigure}[b]{0.35\textwidth}
                \includegraphics[width=\textwidth]{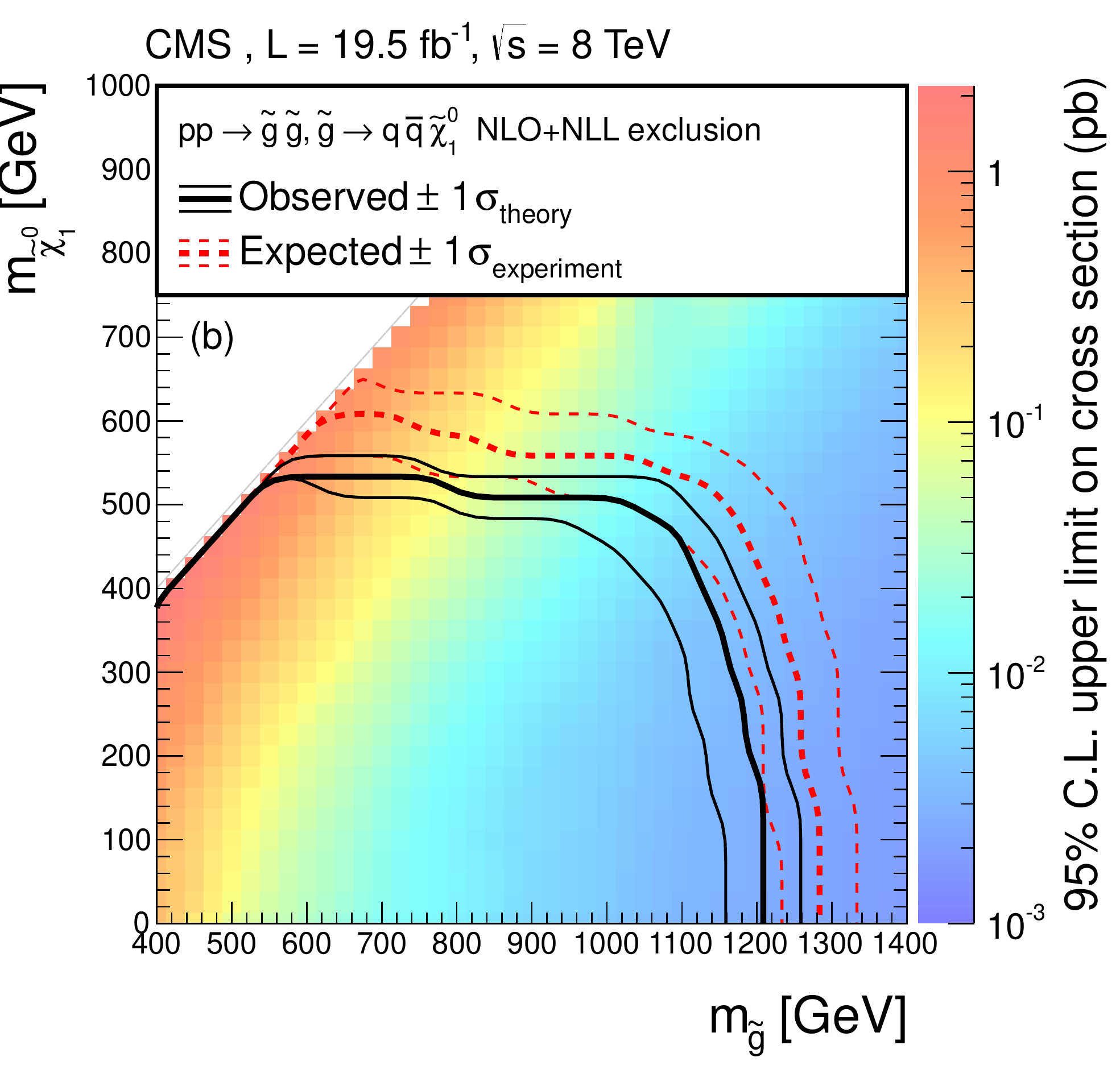}
                \caption{}
                \label{fig:mjet_t1_limit}
        \end{subfigure}
        ~
        \begin{subfigure}[b]{0.35\textwidth}
                \includegraphics[width=\textwidth]{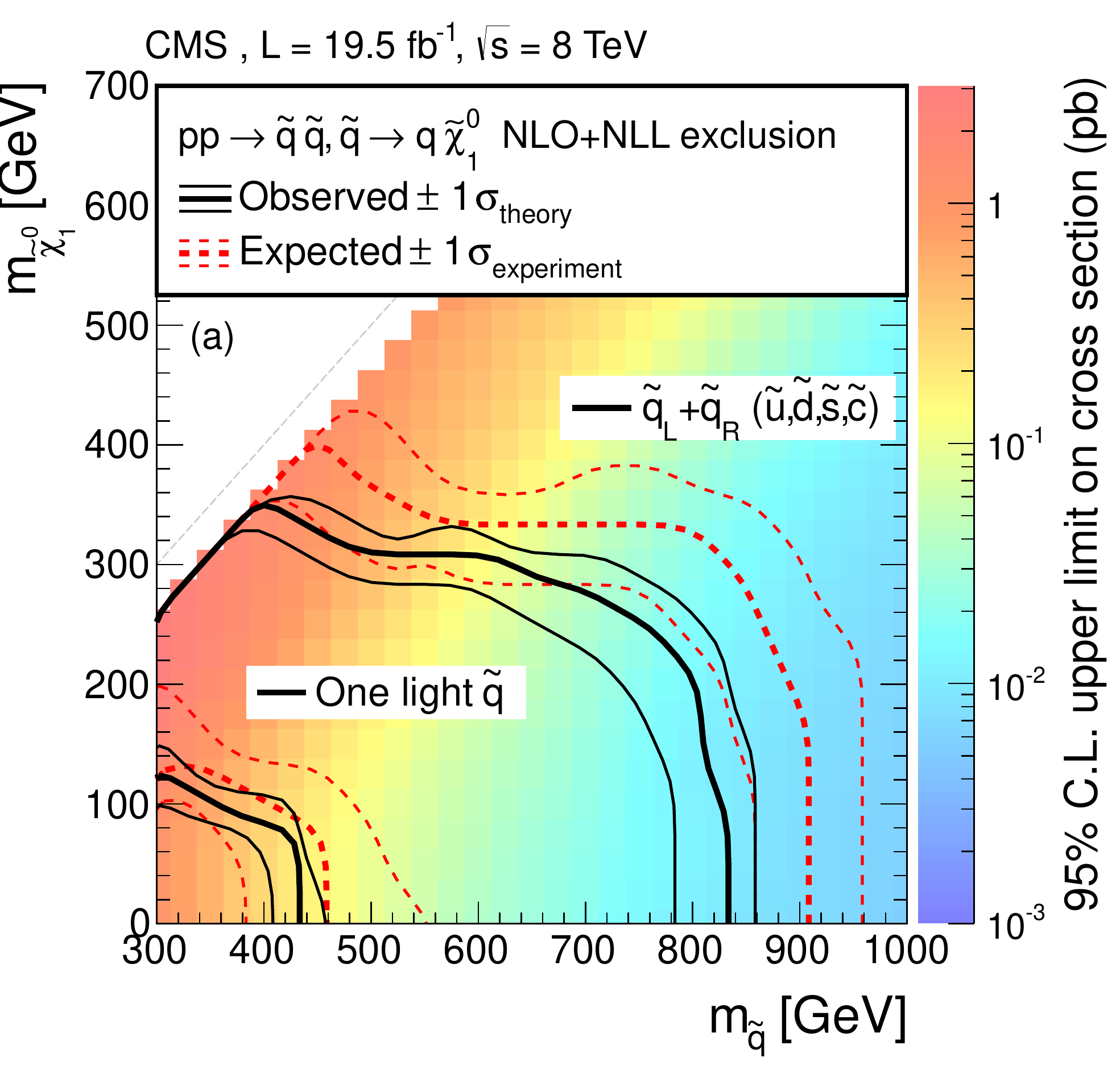}
                \caption{}
                \label{fig:mjet_t2_limit}
        \end{subfigure}
        \caption{Multjiet+\met results interpretted in simplified models of (\ref{fig:mjet_t1_limit}) gluino pair-production, decaying via \tone and (\ref{fig:mjet_t2_limit}) squark pair-production, decaying via \ttwo}
        \label{fig:mjet_limits}
\end{figure}

The analysis categorises events based on the event-variables \Ht, \mht and the jet
multiplicity, with results shown in this binning in Figure
\ref{fig:mjet_results}. No significant excess over the SM background predictions is observed in data and so limits are set in 
models of squark and gluino pair-production, shown in Figure \ref{fig:mjet_limits}.

The limit for light squark pair-production shown in Figure \ref{fig:mjet_t2_limit} has two exclusion curves. The stronger of the two considers a model in which production of all 4 light-squark flavours and their two helicities are available and degenerate. However the weaker of the two represents a model in which this 8-fold degeneracy is removed, and only a single light squark (of an arbitrary flavour and helicity) is produced.

\subsection{MT2 Hadronic}
The analysis described in this section makes use of a novel kinematic variable, 
\mttwo, defined in Equation \ref{eq:mt2_defn} as a function of the LSP pair mass ($m_{\tilde{\chi}}$) \cite{CMS:2014ksa}.

\begin{equation}
\label{eq:mt2_defn}
\ensuremath{
M_{T2}(m_{\tilde{\chi}}) = \underset{\vec{p}_T^{\tilde{\chi}(1)} + \vec{p}_T^{\tilde{\chi}(2)} = \vec{p}_T^{miss}}{min}
\left [ max(M_T^{(1)}, M_T^{(2)}) \right ]
}
\end{equation}

The \mttwo variable can be thought of as a supersymmetric transverse mass, where the masses of the two `invisible' LSP's ($m_{\tilde{\chi}}(1)$ and $m_{\tilde{\chi}}(2)$) are determined from an events kinematic observables. The variable is used here as a discovery variable, where any excesses due to the presence of a supersymmetric decay would be evident in the high-mass tails of an \mttwo distribution.

\begin{figure}[ht]
\centering
\includegraphics[width=.45\textwidth]{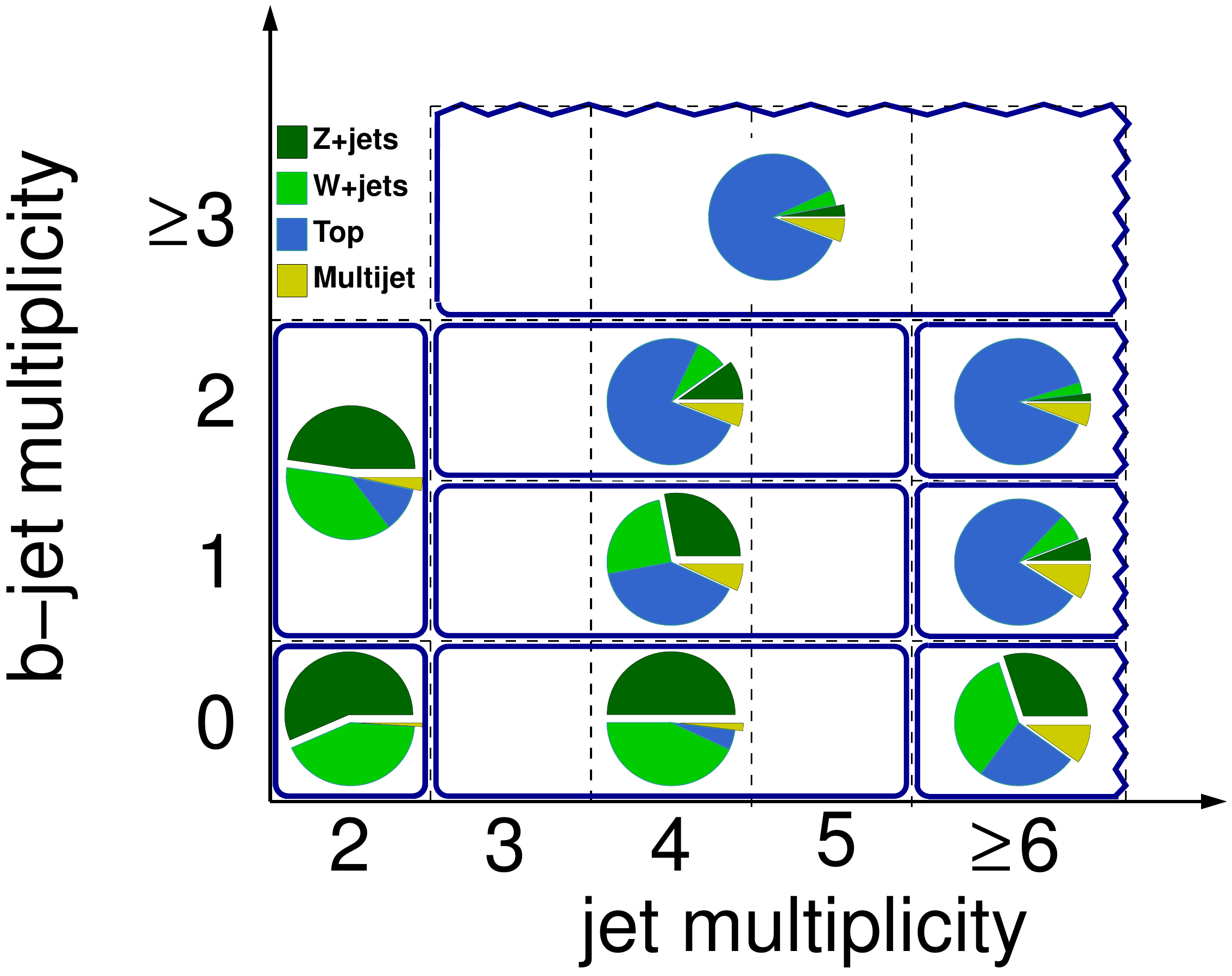}
\caption{Background composition in categories of \njet and \nb for an inclusive \Ht and \mttwo selection.}
\label{fig:mt2_bg_comp}
\end{figure}

The analysis maintains sensitivity to a broad range of hadronic final states by
binning in \mttwo, \Ht, \njet and \nb. The 
\mttwo analysis shares very similar backgrounds to the analysis described in the previous section due to the all-hadronic signal region. In order to estimate these background contributions to the signal region, data-driven techniques are also employed. As demonstrated above in section \ref{sec:multijetmet}, one of the most significant backgrounds to an all-hadronic analysis is QCD multijet -  a background which populates the low \mttwo mass region. To reduce this background, a lower bound on the \mttwo variable is chosen such that the QCD multijet contribution is considered negligible with respect to the total background prediction. This procedure is performed in each of the analysis categories, in dimensions of \Ht, \njet and \nb. Figure \ref{fig:mt2_bg_comp} shows the background composition in \njet and \nb dimensions.

\begin{figure}[h!]
\centering
\includegraphics[width=.6\textwidth]{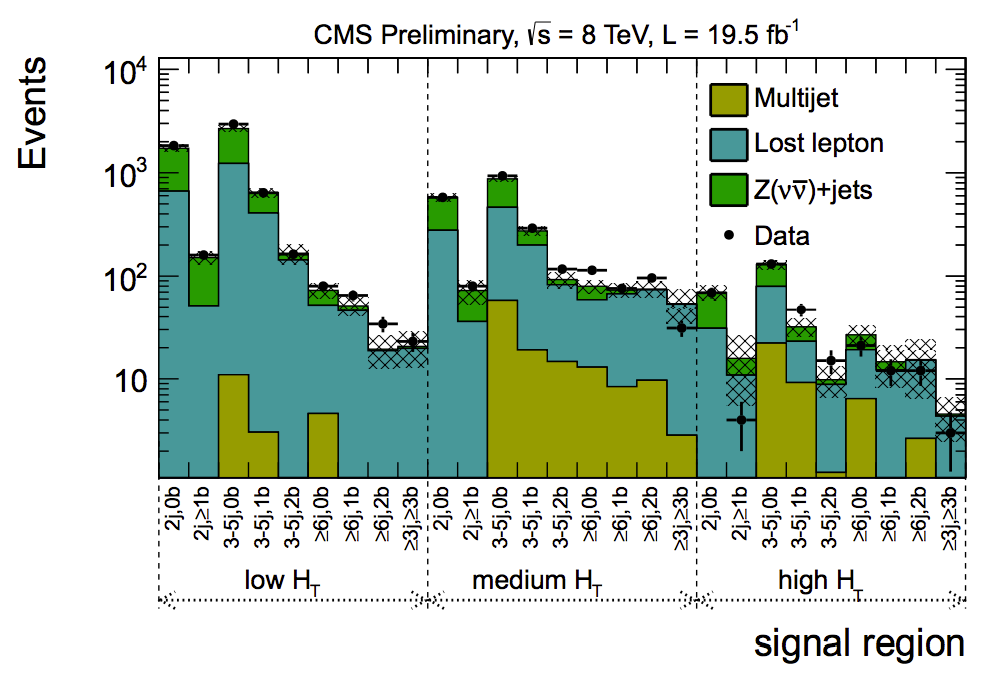}
\caption{Data observations shown for each analysis category of the \mttwo analysis, with the corresponding data-driven Standard Model background estimations.}
\label{fig:mt2_results}
\end{figure}

\begin{figure}[h!]
        \centering
        \begin{subfigure}[b]{0.35\textwidth}
                \includegraphics[width=\textwidth]{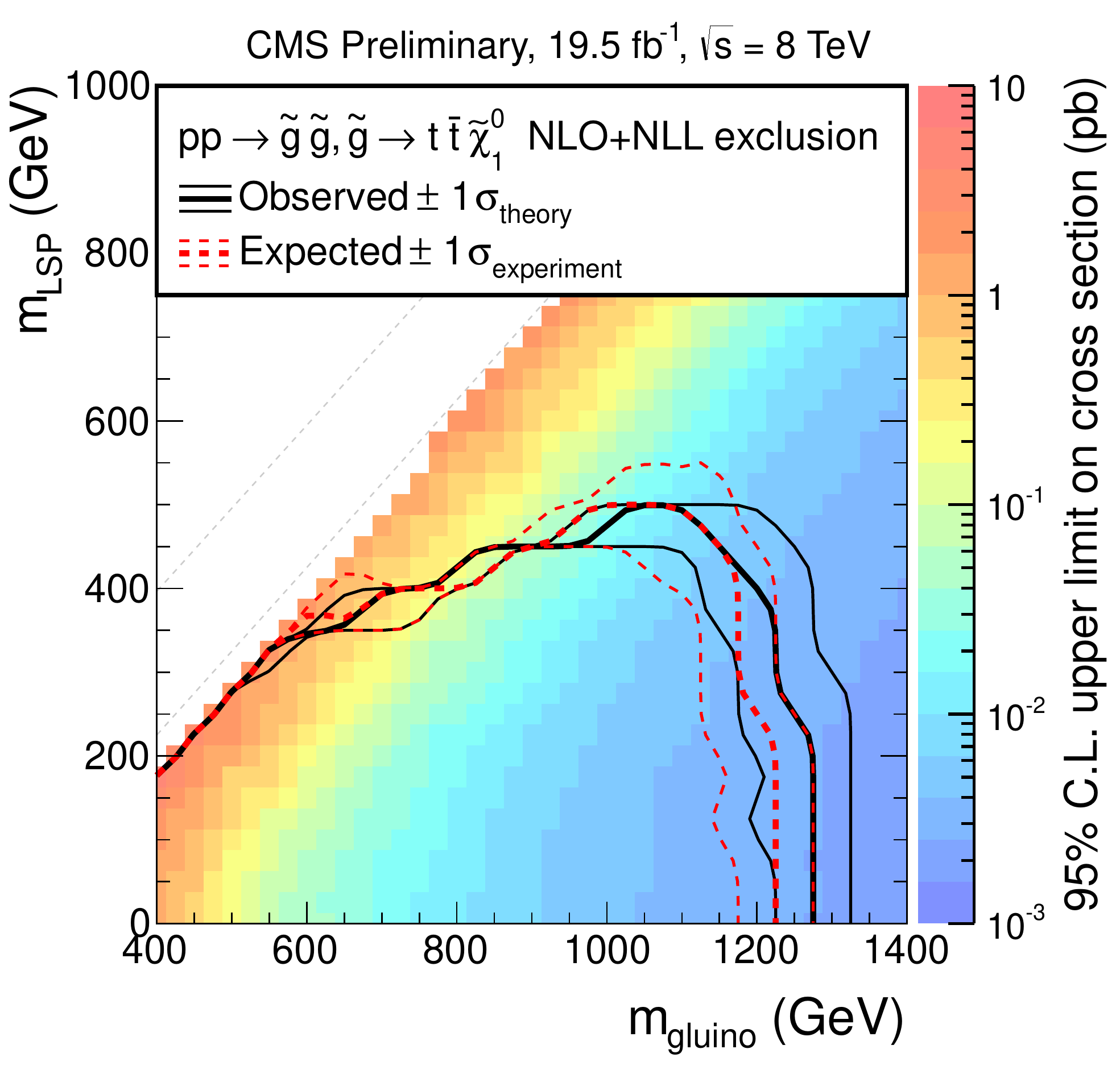}
                \caption{}
                \label{fig:mt2_t1tttt_limit}
        \end{subfigure}
        ~
        \begin{subfigure}[b]{0.35\textwidth}
                \includegraphics[width=\textwidth]{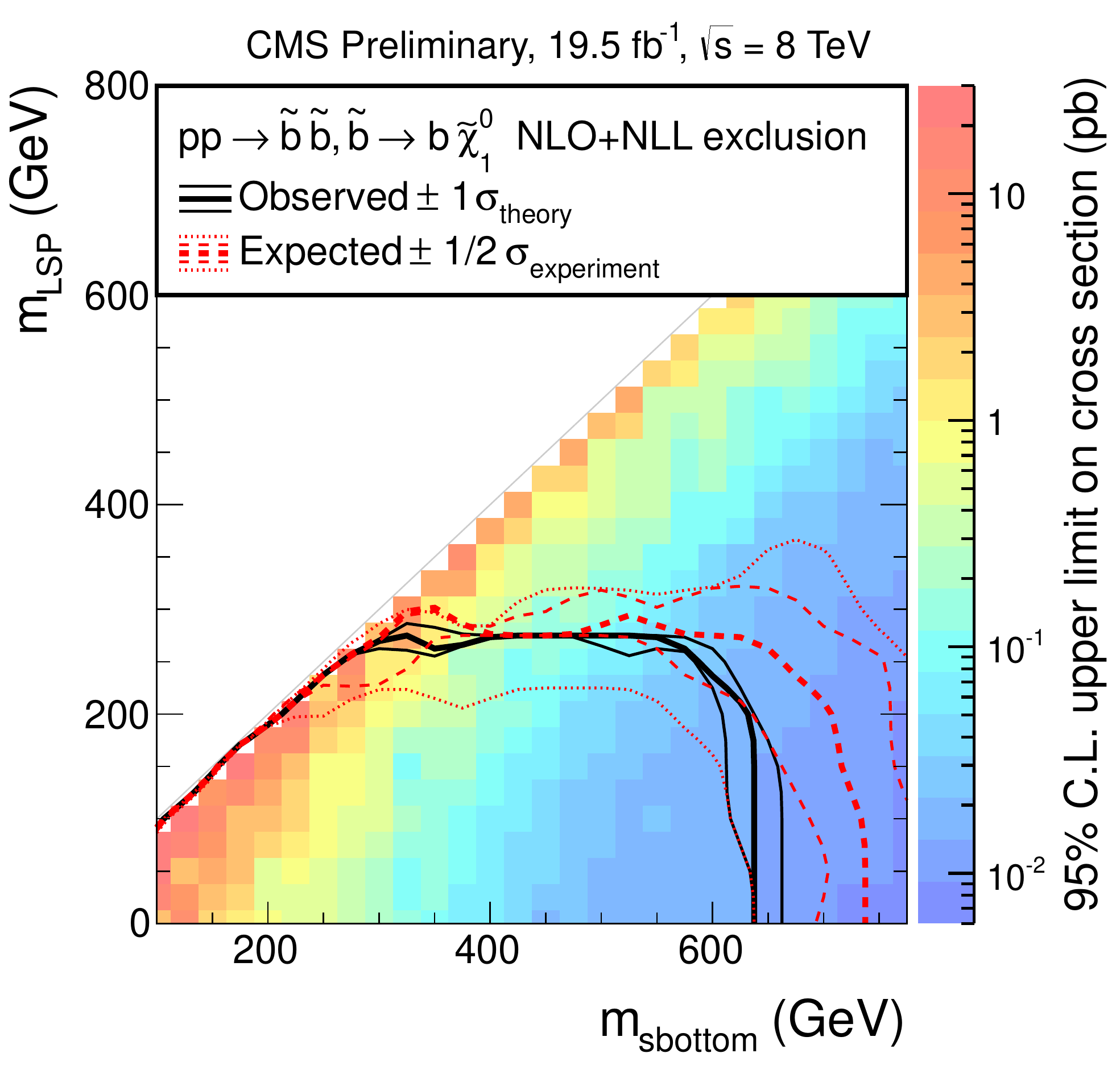}
                \caption{}
                \label{fig:mt2_t2bb_limit}
        \end{subfigure}
        \caption{\mttwo analysis results interpreted in simplified models of (\ref{fig:mt2_t1tttt_limit}) gluino pair-production, decaying via \tonetttt and (\ref{fig:mt2_t2bb_limit}) sbottom pair-production, decaying via \ttwobb..}
        \label{fig:mt2_limits}
\end{figure}

Results are shown in Figure \ref{fig:mt2_results} for the various \Ht regions, subdivided into the multiple \njet and \nb categories. As no significant excess in data is observed over the background prediction, limits are set in multiple models. Targeted interpretations are made using specific event categories with the strongest sensitivities to particular SMS models. Two examples are shown in Figure \ref{fig:mt2_limits} where both squark and gluino production is considered, decaying to heavy quarks and subsequently large numbers of bottom quarks. Naturally high \nb categories are used, increasing signal acceptance while reducing the overall background. This targeted interpretation approach provides very strong limits for such high b-jet multiplicity models, and is also used for a variety of other signal models not shown here \cite{CMS:2014ksa}.

 \subsection{Single lepton with btags}
An inclusive search considers SUSY decays to a single lepton and multiple jets\cite{Chatrchyan:2013iqa}. In such a leptonic final state, dominant standard model backgrounds come from semileptonic \ttbar decays as well as W and Z boson leptonic decays, all estimated using data-driven techniques. The analysis consists of two complementary search strategies, one of which models the \met distribution in exclusive bins of \Ht, while the other relies on the measurement of the azimuthal angle between the mother-particle of the lepton, here assumed to be a standard model W, and it's daughter, \dphiwl. For the latter, under a standard model hypothesis, the kinematic constraints due to the mass of the W boson mother particle lead to an upper-bound on the value of \dphiwl, producing a sharp cutoff in the distribution. However, if the lepton originates from the decay of some supersymmetric particle, the mass of the mother particle can potentially be considerably higher, thereby increasing this upper-bound and populating the tails of a \dphiwl distribution. A cut of \dphiwl $< 1.0$ is made on all events, greatly increasing the signal to background ratio in this signal region.

\begin{figure}[t!]
\centering
	\begin{subfigure}[b]{0.25\textwidth}
		\includegraphics[width=\textwidth]{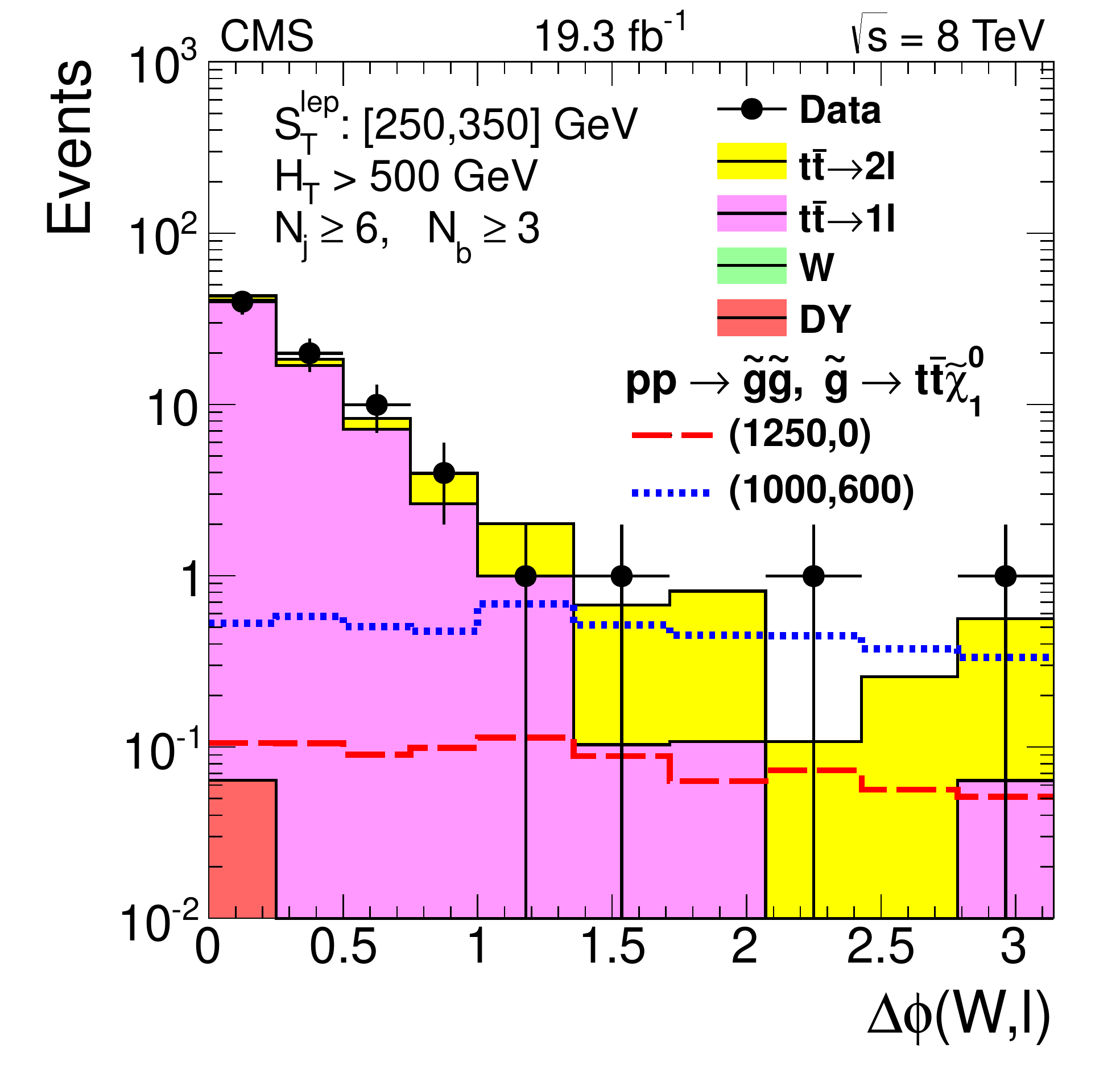}
		\caption{$250 \le \stlep < 350$ \gev}
		\label{fig:sinlep_results_lo}
	\end{subfigure}
	~
	\begin{subfigure}[b]{0.25\textwidth}
		\includegraphics[width=\textwidth]{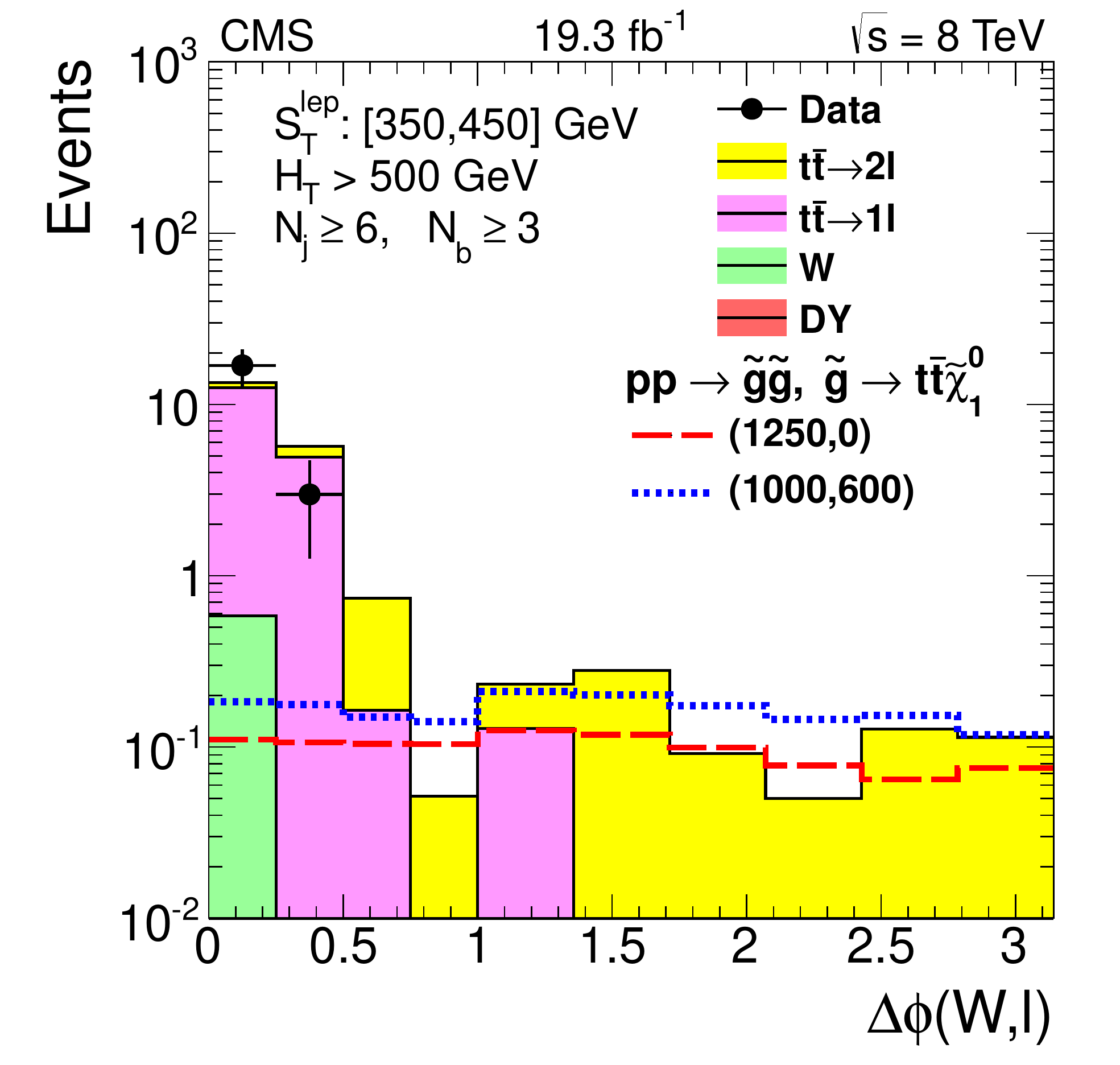}
		\caption{$350 \le \stlep < 450$ \gev}
		\label{fig:sinlep_results_mid}
	\end{subfigure}
	~
	\begin{subfigure}[b]{0.25\textwidth}
		\includegraphics[width=\textwidth]{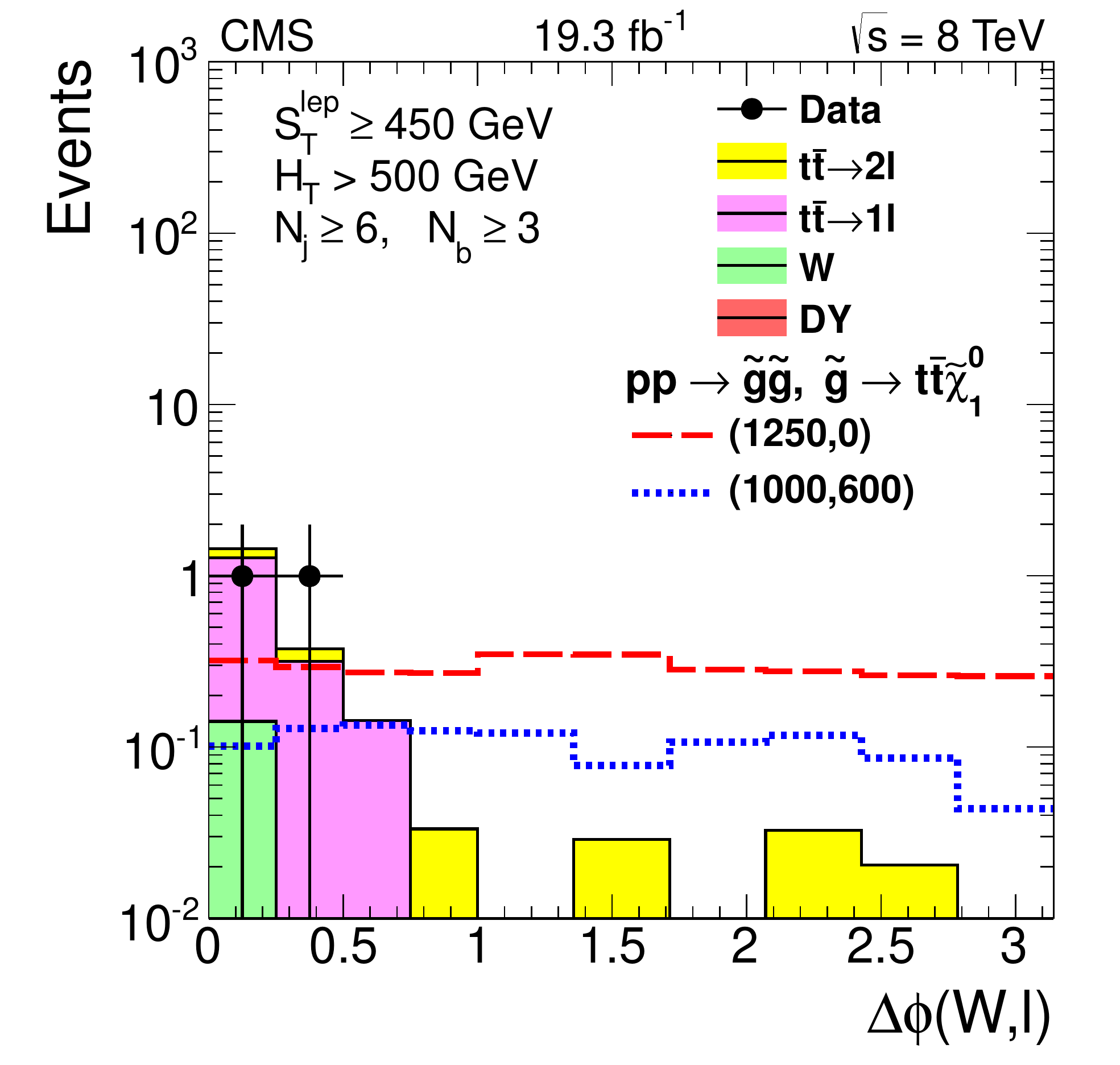}
		\caption{$\stlep \ge 450$ \gev}
		\label{fig:sinlep_results_hi}
	\end{subfigure}
\caption{Results for the single lepton + b analysis, showing the \dphiwl distributions for different \stlep bins in the $\Ht >500$\gev, $\njet \ge 6$ and $\nb \ge 3$ category.}
\label{fig:sinlep_results}
\end{figure}

The analysis is binned in \njet and \nb, as well as the scalar sum of transverse leptonic energy in the event, \stlep. Results for a given \njet, \b and \Ht category with the \dphiwl method are shown in Figure \ref{fig:sinlep_results}.


\begin{figure}[ht]
\centering
	\begin{subfigure}[b]{0.35\textwidth}
		\includegraphics[width=\textwidth]{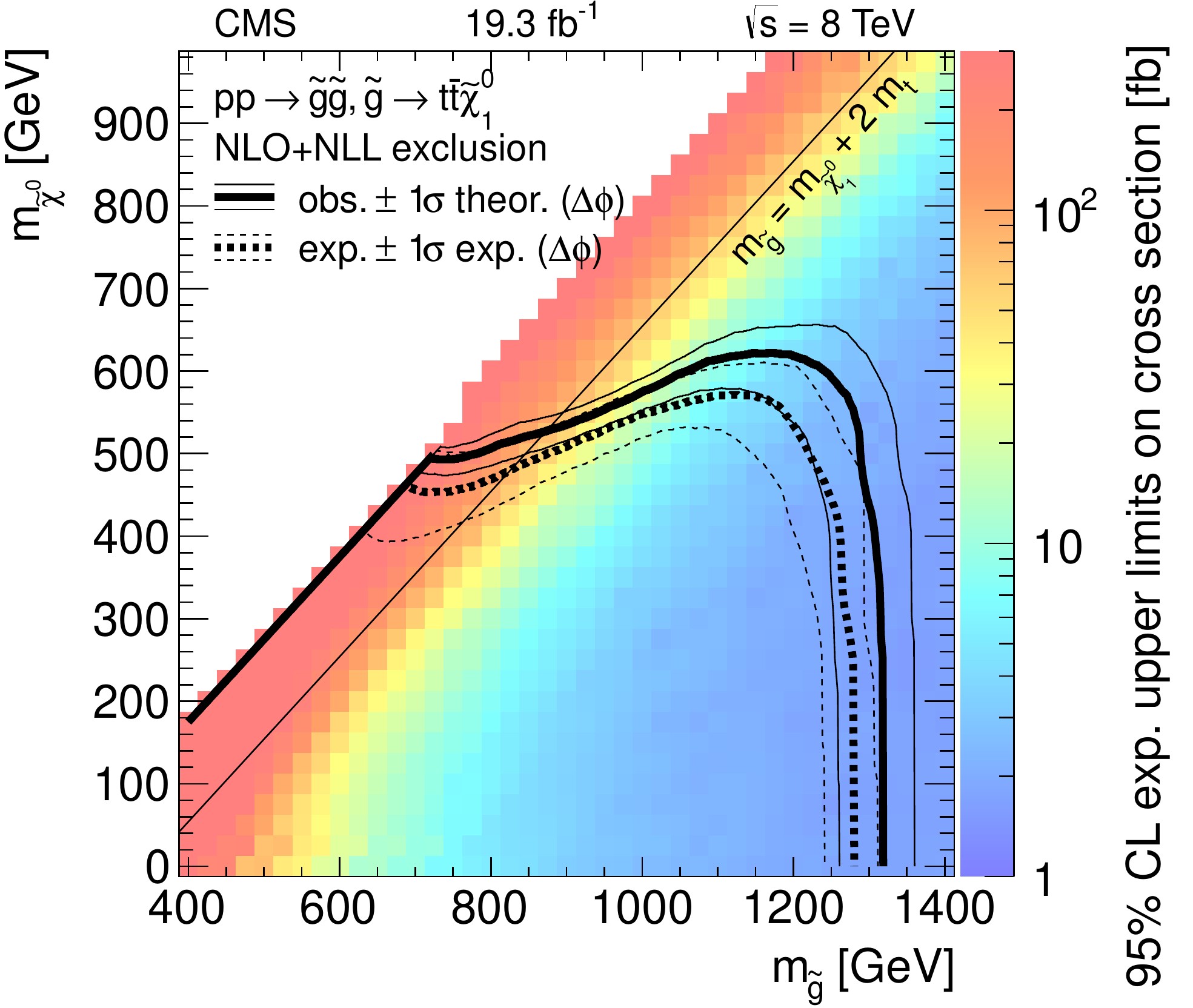}
		\caption{}
		\label{fig:sinlep_t1tttt_limit}
	\end{subfigure}
	~
	\begin{subfigure}[b]{0.35\textwidth}
		\includegraphics[width=\textwidth]{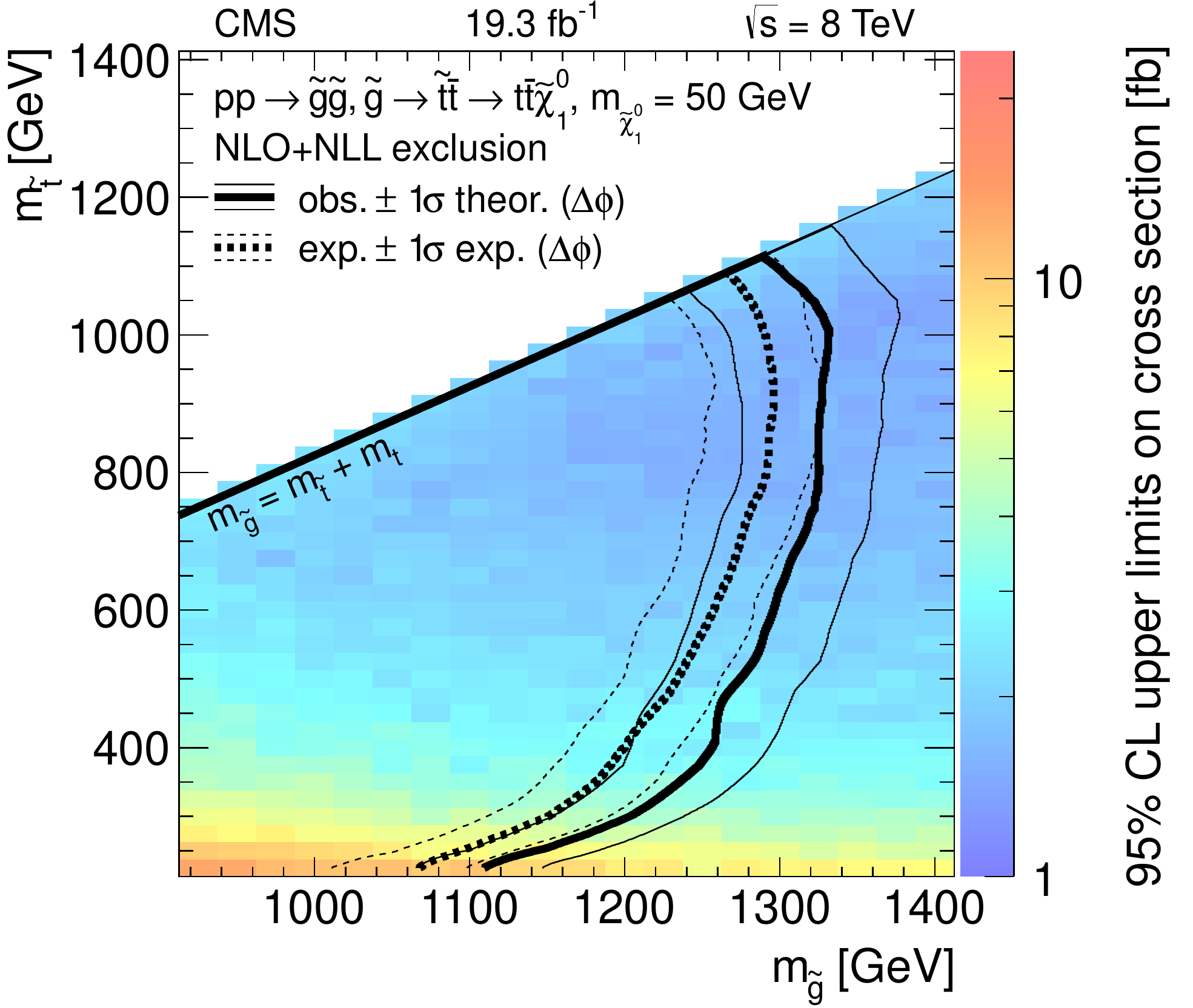}
		\caption{}
		\label{fig:sinlep_t5tttt_limit}
	\end{subfigure}
\caption{Limit from the single lepton + \met search for the (\ref{fig:sinlep_t1tttt_limit}) simplified model of gluino production, with the decay \tonetttt, and (\ref{fig:sinlep_t5tttt_limit}) gluino production with the decay \tonetonet.}
\label{fig:sinlep_limit}
\end{figure}

No statistically significant excess is observed and so limits are placed in multiple models, two of which is shown in Figure \ref{fig:sinlep_limit}. Strong exclusions in $m_{\tilde{g}}$, $m_{\stop1}$ and $m_{\tilde{\chi}_0}$ are achieved due to the strong background discriminating power of the \dphiwl method.

\subsection{Tri-lepton with btags}
\label{sec:trilep}

The final analysis considered here requires at least 3 leptons with associated b-tagged jets and \met \cite{Khachatryan:2014doa}. Although requiring such a high number of leptons reduces the SUSY production branch fraction, the effect is greatly offset by strongly suppressing any standard model backgrounds. Following these requirements, the only remaining backgrounds come from events with three prompt leptons, for example from WZ di-boson production, and a small contribution from rare standard model processes such as $t\bar{t}+V/H$, $t\bar{b}+Z$ and $VVV$ (where V represents a vector boson). While the latter contribution is very small, the former is suppressed by the requirement of b-tagged jets.

\begin{figure}
        \centering
        \begin{subfigure}[b]{0.7\textwidth}
                \includegraphics[width=\textwidth]{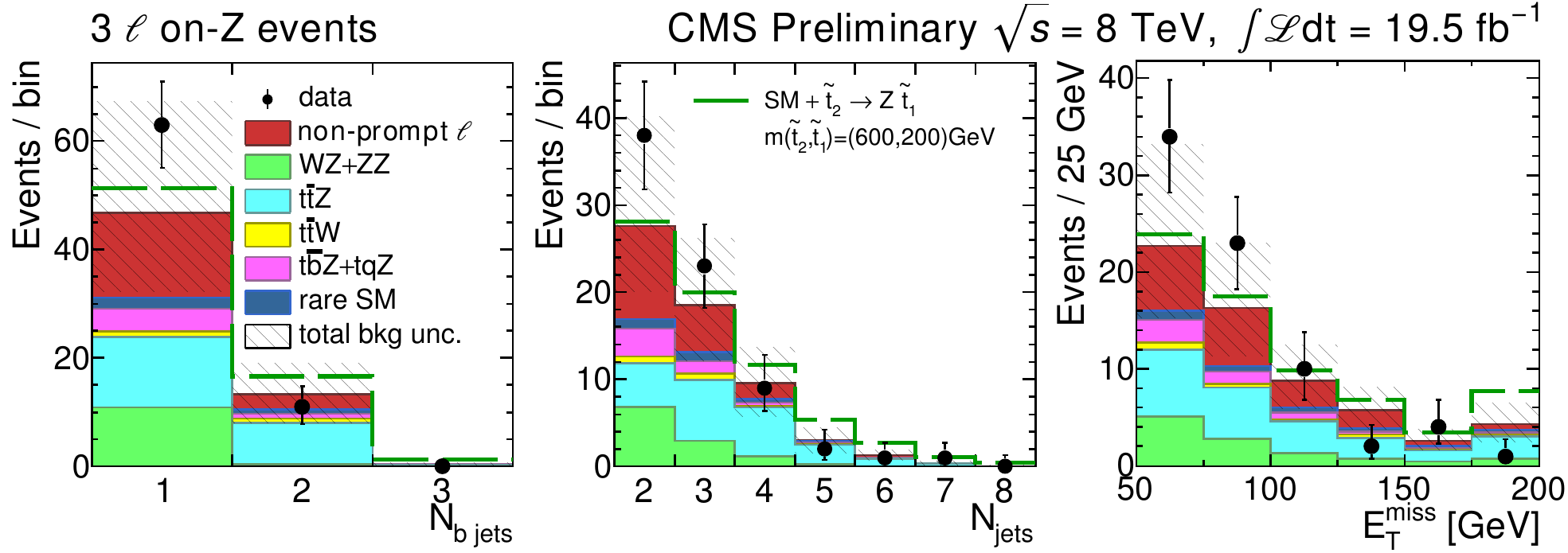}
                \caption{On-Z}
                \label{fig:trilep_results_onZ}
        \end{subfigure}
        \\
        \begin{subfigure}[b]{0.7\textwidth}
                \includegraphics[width=\textwidth]{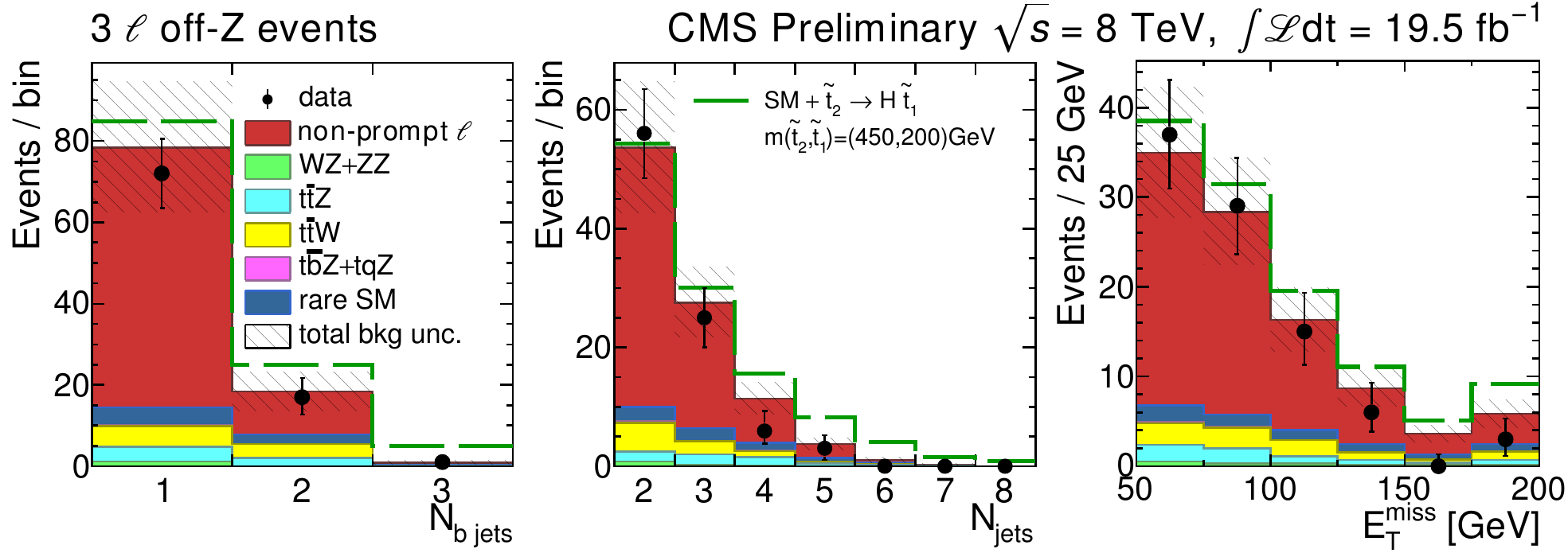}
                \caption{Off-Z}
                \label{fig:trilep_results_offZ}
        \end{subfigure}
        \caption{Results from the trilepton + b analysis, showing (\ref{fig:trilep_results_onZ}) the on-Z mass and (\ref{fig:trilep_results_offZ}) the off-Z mass regions.}
        \label{fig:trilep_results}
\end{figure}

The search is carried out in bins of \Ht, \met, \njet and \nb, allowing the analysis to remain inclusive to a wide number of possible signal scenarios through targeted interpretation. By calculating the invariant mass of lepton pairs in the tri-leptonic system, two regions are formed as either `on' or `off' the Z-mass pole, with each region having different background compositions. Results for this binning schema are shown in Figure \ref{fig:trilep_results}.

\begin{figure}[ht]
        \centering
        \begin{subfigure}[b]{0.2\textwidth}
                \includegraphics[width=\textwidth]{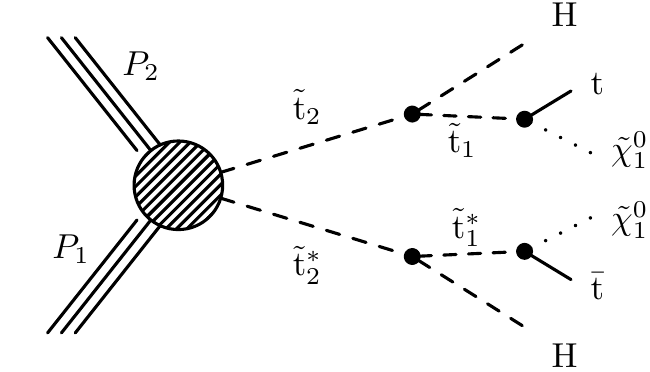}
                \caption{0\% Z}
                \label{fig:trilep_feyn_zero}
        \end{subfigure}
        ~
        \begin{subfigure}[b]{0.2\textwidth}
                \includegraphics[width=\textwidth]{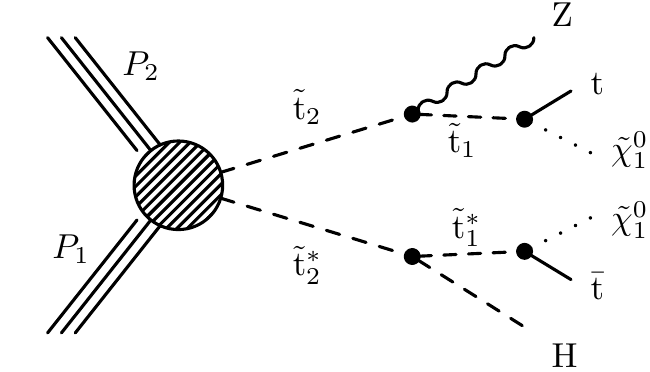}
                \caption{50\% Z}
                \label{fig:trilep_feyn_fifty}
        \end{subfigure}
        ~
        \begin{subfigure}[b]{0.2\textwidth}
                \includegraphics[width=\textwidth]{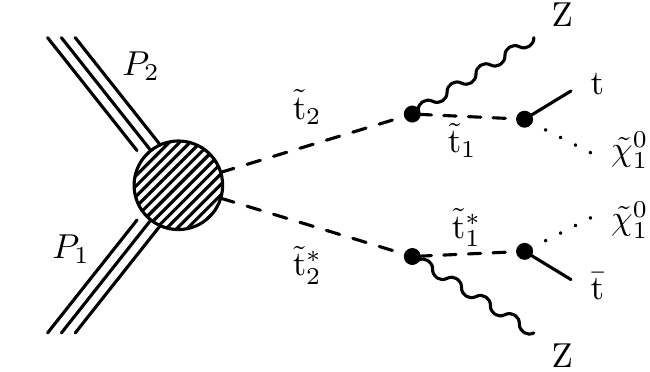}
                \caption{100\% Z}
                \label{fig:trilep_feyn_hundred}
        \end{subfigure}
        \caption{Model decays used for interpretation of the tri-lepton + b analysis.}
        \label{fig:trilep_feyn}
\end{figure}
 
\begin{figure}[h!]
\centering
	\begin{subfigure}[b]{0.35\textwidth}
		\includegraphics[width=\textwidth]{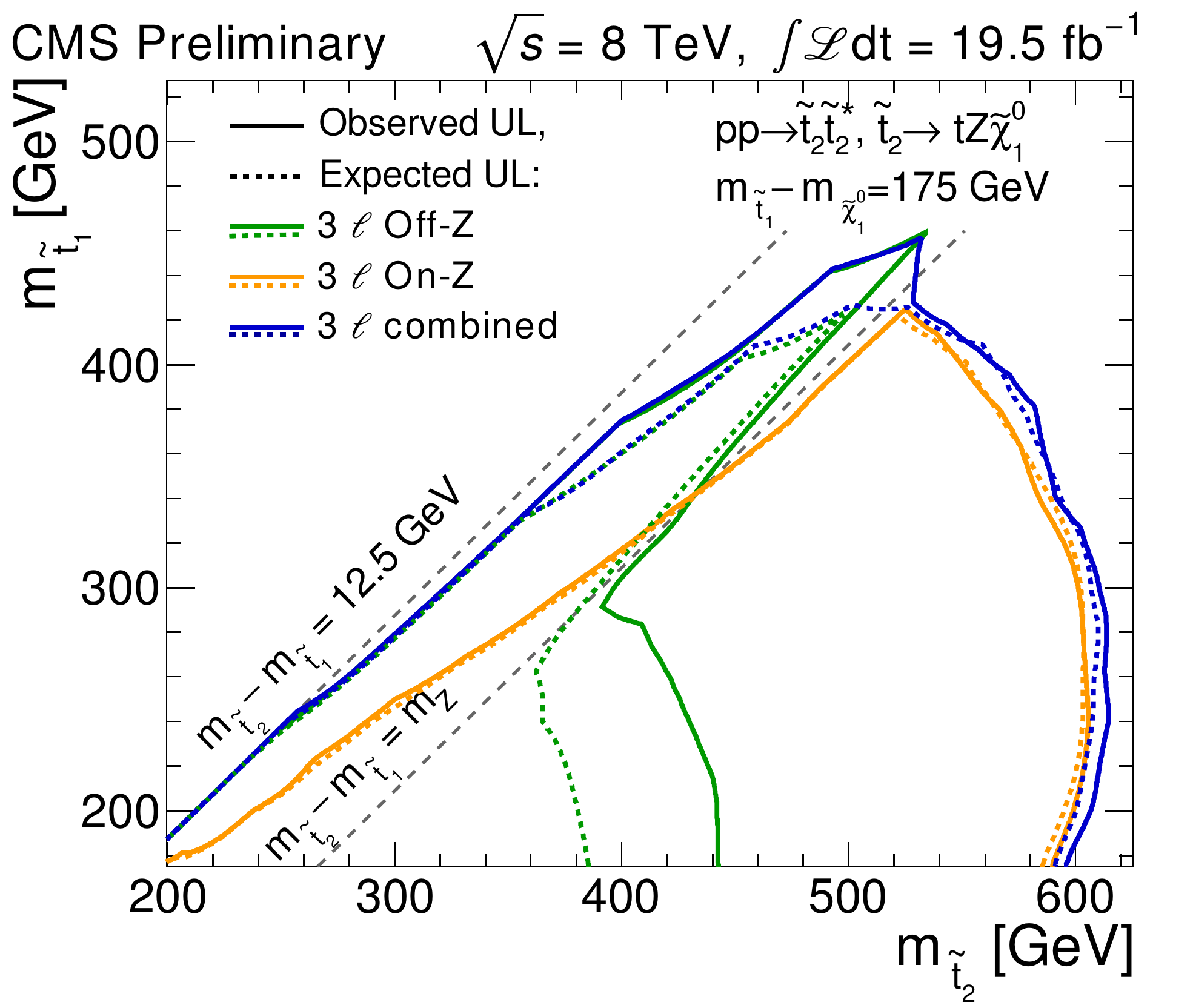}
		\caption{}
		\label{fig:trilep_limit_hundredZ}
	\end{subfigure}
	~
	\begin{subfigure}[b]{0.335\textwidth}
		\includegraphics[width=\textwidth]{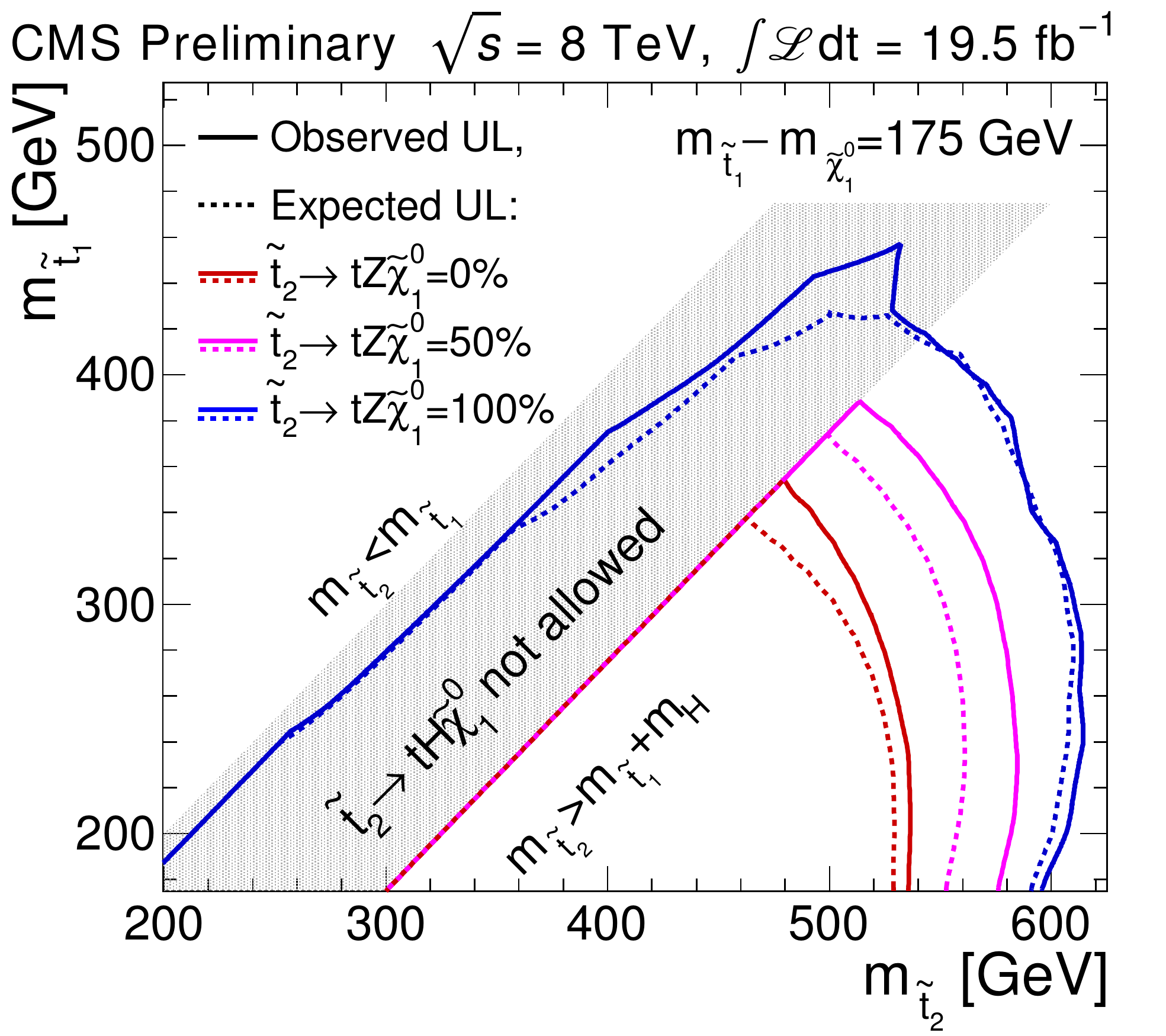}
		\caption{}
		\label{fig:trilep_limit_all}
	\end{subfigure}
\caption{Limits from the trilepton + b analysis on the 100\% scenario (\ref{fig:trilep_limit_hundredZ}), showing limits from the on and off Z-mass regions separately and combined, and for different admixtures of the $\tilde{t}_2$ decays via Z and H (\ref{fig:trilep_limit_all}).}
\label{fig:trilep_limit}
\end{figure}

As no significant excesses are observed, interpretations are made in a variety of models. Of these, one of particular interest is that of $\tilde{t}_2$ pair production, where the $\tilde{t}_2$ decays via the lighter $\tilde{t}_1$ with either $\tilde{t}_2 \rightarrow \tilde{t}_1 + Z$ or $\tilde{t}_2 \rightarrow \tilde{t}_1 + H$, followed by $\tilde{t}_1 \rightarrow t + \neu1$, when $m_{\tilde{t}_1} - m_{\neu1} \approx m_t$. This gives rise to the three scenarios shown in Figure \ref{fig:trilep_feyn}, referred to by the relative branching fraction to Z-bosons in the decay. Figure \ref{fig:trilep_limit_hundredZ} shows the limit from each Z-mass region for the 100\% Z scenario (Figure \ref{fig:trilep_feyn_hundred}).  As expected, the off-Z mass category drives the limit when the mass splitting between \stop1 and \stop2 becomes less than the mass of the Z, with the opposite being true for the on-Z category. Limits for all three decay scenarios are shown in Figure \ref{fig:trilep_limit_all}, with the 100\% Z limit being the strongest.

\section{Conclusions}

A selection of inclusive CMS SUSY searches were presented, using the full 19.7\fb dataset collected in 2012. These included both leptonic and hadronic analyses, and are part of a much larger SUSY program which aims to cover the largest area of SUSY phase space possible with other inclusive and targeted searches. As no statistically significant signal has been observed in the current dataset, strong limits have been produced, for example gluino masses in excess of 1 \tev and stop masses up to 600 \gev dependent on the decay mode. However, with the increase of \sqs in Run \RNum{2}, a significant gain in parton luminosity will be realised. As an example, the relative production cross section for a 1.5 \tev gluino will increase by a factor of 36 with respect to Run \RNum{1}. Despite the stringent limits presented here, significant amounts of SUSY phase space remain un-probed and so we look towards 2015 for a glimpse of supersymmetry at the LHC.

\Acknowledgements
My sincere thanks go to the LHCP organisers for arranging a well-planned and fruitful 
conference, my CMS colleagues for running and delivering such high-quality data, 
in particular those that produced these competitive results, and my PhD 
supervisor for giving me the opportunity to attend this conference, eat too much of 
New York's food and a drink a large amount of it's beer.

\end{document}